\documentclass[11pt,preprint]{elsarticle}

\usepackage{fancyhdr}
\usepackage{fullpage}
\usepackage{amsmath}
\usepackage{amsbsy}
\usepackage{amssymb}
\usepackage{amscd}
\usepackage{amsfonts}
\usepackage{supertabular}
\usepackage{graphics}
\usepackage{verbatim}
\usepackage{epsfig}
\usepackage{xspace}
\usepackage{euscript}
\usepackage{alltt}
\usepackage{boxedminipage}
\usepackage{float}
\usepackage[colorlinks]{hyperref}
\usepackage{color}
\usepackage[all]{xy}
\usepackage{t1enc}
\usepackage{times,exscale}
\usepackage{graphicx,calc}
\usepackage{subfig}
\usepackage[ruled,vlined]{algorithm2e}
\usepackage{epstopdf}
\usepackage{multirow}
\usepackage{pseudocode}
\usepackage{braket}
\usepackage{esint}

\def\bR{{\mathbf{R}}}
\def\bx{{\mathbf{x}}}
\def\bL{{\mathbf{L}}}

\def\bk{{\mathbf{k}}}
\def\R{{\mathbb{R}}}
\def\C{{\mathbb{C}}}

\def\bA{\textbf{A}}
\def\bB{\textbf{B}}
\def\bC{\textbf{C}}
\def\bX{{\mathbf{X}}}
\def\bw{\mathbf{\bigwedge}}

\def\bxi{{\boldsymbol{\xi}}}
\def\benta{{\boldsymbol{\eta}}}
\def\bzeta{{\boldsymbol{\zeta}}}
\def\bDchis{{\mathbf{D}_{\xi\xi}}}
\def\bDetas{{\mathbf{D}_{\eta\eta}}}
\def\bDetasT{{\mathbf{D}_{\eta\eta}^T}}
\def\bDzetas{{\mathbf{D}_{\zeta\zeta}}}
\def\bDzetasT{{\mathbf{D}_{\zeta\zeta}^T}}
\def\bDchi{{\mathbf{D}_\xi}}
\def\bDeta{{\mathbf{D}_\eta}}
\def\bDetaT{{\mathbf{D}_{\eta}^T}}
\def\bDzeta{{\mathbf{D}_\zeta}}
\def\bDzetaT{{\mathbf{D}_{\zeta}^T}}
\def\bIa{{\mathbf{I}_{n_1}}}
\def\bIb{{\mathbf{I}_{n_2}}}
\def\bIbT{{\mathbf{I}_{n_2}^T}}
\def\bIc{{\mathbf{I}_{n_3}}}
\def\bIcT{{\mathbf{I}_{n_3}^T}}

\journal{arXiv}

\begin{document}

\begin{frontmatter}
\title{On real-space Density Functional Theory for non-orthogonal crystal systems: Kronecker product formulation of the kinetic energy operator}
\author[gatech]{Abhiraj Sharma}
\author[gatech]{Phanish Suryanarayana\corref{cor}}
\address[gatech]{College of Engineering, Georgia Institute of Technology, GA 30332, USA}
\cortext[cor]{Corresponding Author (\it phanish.suryanarayana@ce.gatech.edu) }


\begin{abstract}
We present an accurate and efficient real-space Density Functional Theory (DFT) framework for the ab-initio study of non-orthogonal crystal systems. Specifically, employing a local reformulation of the electrostatics, we develop a novel Kronecker product formulation of the real-space kinetic energy operator that significantly reduces the number of operations associated with the Laplacian-vector multiplication, the dominant cost in practical computations. In particular, we reduce the scaling with respect to finite-difference order from quadratic to linear, thereby significantly bridging the gap in computational cost between non-orthogonal and orthogonal systems. We verify the accuracy and efficiency of the proposed methodology through selected examples. 
\end{abstract}

\end{frontmatter}
\section{Introduction}
The high accuracy to cost ratio of Kohn-Sham Density Functional Theory (DFT) \cite{Hohenberg,Kohn1965} as an ab-initio method makes it a very attractive tool for understanding and predicting a wide range of material properties. Among the various DFT implementations, the plane-wave basis has been a particularly popular choice for the discretization of the Kohn-Sham equations \cite{VASP,CASTEP,ABINIT,Espresso,DFT++,gygi2008architecture}. This is motivated by a number of attractive features of the plane-wave method, including the ability to study various crystal systems with differing symmetry at similar computational cost. However, the plane-wave basis suffers from a number of limitations, including its nonlocal nature, which can limit its efficiency. This is particularly the case in the context of scalable high-performance computing. 

In order to overcome the limitations of plane-waves, there have been a number of efforts directed towards the development of real-space approaches for DFT, wherein the equations are discretized using high-order central finite-differences \cite{chelikowsky1994finite,OCTOPUS,shimojo2001linear,Iwata2010,ghosh2017sparc,ghosh2017extended}. Notably, these approaches are highly competitive with their plane-wave counterparts, achieving speedups of up to an order of magnitude in some cases \cite{ghosh2017sparc,ghosh2017extended}. However, the cost of the Laplacian-vector multiplication---key computational kernel that determines the efficiency of real-space DFT calculations---is significantly larger for non-orthogonal systems compared to the analogous orthogonal ones. This is due to the mixed derivatives arising in the Laplacian for non-orthogonal coordinate axes, which makes its product with a vector scale quadratically with respect to the finite-difference order, unlike the linear scaling for orthogonal systems. Since commonly employed discretization orders can be as large as twelve, real-space DFT calculations for non-orthogonal systems are significantly more expensive than their orthogonal counterparts.

An alternative to central finite-differences in real-space DFT is the use of Mehrstellan methods \cite{briggs1996real,fattebert1999finite,fattebert2000}---an expansion technique that utilizes more local information. However, such a discretization varies with the type of non-orthogonal grid and results in a non-Hermitian generalized eigenvalue problem, which can limit the efficiency of the ensuing calculations. In view of the aforementioned limitations of real-space methods for non-orthogonal systems, a new technique was proposed in which additional directions are introduced into the Laplacian in order to remove the mixed derivative terms \cite{natan2008real}. In this approach, although the computational cost associated with the Laplacian-vector multiplication is significantly reduced and the scaling is linear with respect to the finite-difference order, the effective grid spacing for the new directions can be larger than that in the lattice vector directions. Consequently, finer meshes might be required than those for analogous orthogonal systems, thereby limiting the efficiency of such a strategy. 

In this work, we present an accurate and efficient framework for performing real-space Density Functional Theory (DFT) calculations of non-orthogonal crystal systems. Specifically, while utilizing a local reformulation of the electrostatics \cite{pask2005, Phanish2010} that is equally applicable to systems with different crystal symmetries, we develop a new Kronecker product \cite{van2000ubiquitous,golub2012matrix} formulation of the real-space kinetic energy operator that significantly reduces the operation count associated with the Laplacian-vector multiplication, the dominant cost in real-space DFT computations for small to moderately sized systems ($\sim 1000$ atoms). In particular, we reduce the scaling with respect to the finite-difference order from quadratic to linear, thereby enabling the study of the different crystal systems at similar cost. We verify the accuracy and efficiency of the proposed methodology with selected examples, including a system with the most general crystal symmetry, i.e., triclinic. 

The remainder of this paper is organized as follows. First, we review the real-space formulation of DFT in Section~\ref{Sec:DFT}, followed by the Kronecker product formulation for the kinetic energy operator in Section~\ref{section:kroneckerproduct}. Next, we study the accuracy and efficiency of the proposed framework in Section~\ref{Sec:ImplementationResults}. Finally, we provide concluding remarks in Section~\ref{Sec:conclusions}.

\section{Real-space formulation of Density Functional Theory} \label{Sec:DFT}

\begin{figure}[H]
\centering
\includegraphics[keepaspectratio=true,width=0.35\textwidth]{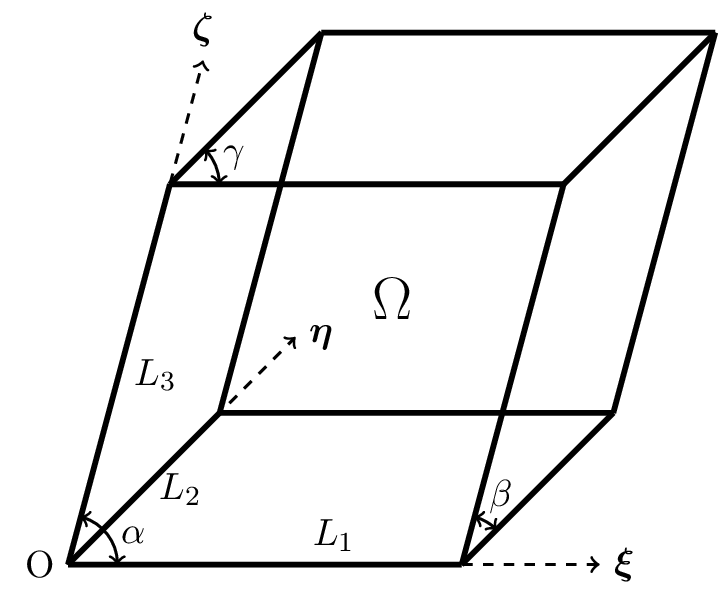}
\caption{Non-orthogonal unit cell $\Omega$ with associated lattice vectors $L_1 \hat{\bxi}$, $L_2 \hat{\benta}$, and $L_3 \hat{\bzeta} $, where $\hat{\boldsymbol{\xi}}$, $\hat{\boldsymbol{\eta}}$, and $\hat{\boldsymbol{\zeta}}$ are the lattice unit vectors. The angles between the lattice unit vectors are: $\alpha = \arccos( \hat{\bxi} \cdot \hat{\bzeta})$, $\beta = \arccos(\hat{\bzeta} \cdot \hat{\benta})$, and $\gamma = \arccos(\hat{\benta} \cdot \hat{\bxi})$.}
\label{Fig:unitcell}
\end{figure}

In this section, we present a formulation of Density Functional Theory (DFT) that is particularly suitable for real-space calculations. Though its accuracy and efficiency has previously been verified for orthogonal systems \cite{ghosh2017extended}, it is equally applicable and effective for non-orthogonal crystal systems, as we show in this work. Consider a non-orthogonal unit cell $\Omega$ with associated lattice vectors $L_1 \hat{\bxi}$, $L_2 \hat{\benta}$, and $L_3 \hat{\bzeta}$, where $\hat{\boldsymbol{\xi}}$, $\hat{\boldsymbol{\eta}}$, and $\hat{\boldsymbol{\zeta}}$ are the lattice unit vectors (Fig.~\ref{Fig:unitcell}). Let the nuclei be positioned at $\bR = \{\bR_1, \bR_2, \ldots, \bR_N \}$, with a total of $N_e$ valence electrons in the system. Neglecting spin, the nonlinear eigenproblem for the electronic ground-state in Kohn-Sham DFT can be written as 
\begin{equation}\label{eqn:eigenvalue}
\left( \mathcal{H}\equiv -\frac{1}{2} \nabla^2 + V_{xc} + \phi + V_{nl} \right)  \psi_n  =  \lambda_n \psi_n \,, \quad n=1,2, \ldots, N_s \,
\end{equation}
where $\mathcal{H}$ is the Hamiltonian, $\psi_n$ are the orbitals with energies $\lambda_n$, $V_{xc}$ is the exchange-correlation potential, $\phi$ is the electrostatic potential \cite{pask2005, Phanish2010}, $V_{nl}$ is the nonlocal pseudopotential operator, and $N_s$ is the number of states. The Kohn-Sham orbitals $\psi_n$ are Bloch-periodic, i.e., for every lattice vector $\bL$ and Bloch wavevector $\bk$, 
\begin{equation}\label{eqn:BlochPBC}
\psi_n(\bx + \bL,\bk) = e^{i \bk . \bL} \psi_n(\bx,\bk) \,.
\end{equation}
The nonlocal pseudopotential operator in Kleinman-Bylander form \cite{kleinman1982efficacious} takes the form
\begin{eqnarray}\label{eqn:nonlocaloperator}
V_{nl} \psi_n = \sum_{I} \sum_{lm} \gamma_{Il} \tilde{\chi}_{Ilm} \left(\int_{\Omega} \tilde{\chi}_{Ilm}^*(\bx,\bR_{I}) \psi_n\, \mathrm{d\bx} \right) \,
\end{eqnarray}
where the summation index $I$ runs over all atoms in $\Omega$, the summation index $lm$ runs over all azimuthal and magnetic quantum numbers, and $\tilde{\chi}_{I'lm}$ are the Bloch-periodically mapped projectors. The electrostatic potential $\phi$ is periodic and satisfies the Poisson equation \cite{pask2005, Phanish2010}:
\begin{eqnarray}\label{eqn:poisson}
-\frac{1}{4\pi} \nabla^2 \phi(\bx,\bR) = \rho(\bx) + b(\bx,\bR) \,,
\end{eqnarray}
where $b$ is the total pseudocharge density and 
\begin{eqnarray}\label{eqn:density}
\rho(\bx) = 2 \sum_{n=1}^{N_s} \fint_{BZ} g_n(\bk) | \psi_n(\bx,\bk) |^2 \, \mathrm{d\bk} 
\end{eqnarray}
is the electron density. Above, $\fint_{BZ}$ represents the volume average of the Brillouin zone and $g_n(\bk)$ are the orbital occupations:
\begin{eqnarray}\label{eqn:occupation}
g_n(\bk) = \Bigg( 1 + \exp\left( \frac{\lambda_n(\bk) - \lambda_f}{\sigma} \right) \Bigg)^{-1} \,, \quad \lambda_f \,\,\, \text{is} \,\,\, s.t. \,\,\, 2 \sum_{n=1}^{N_s} \fint_{BZ} g_n(\bk) \, \mathrm{d \bk} = N_e \,,
\end{eqnarray}
where $\lambda_f$ is the Fermi energy and $\sigma$ is the smearing. 

Once the electronic ground-state has been determined, the free energy can be written as \cite{ghosh2017extended} 
\begin{eqnarray} \label{eqn:freeenergy}
\mathcal{F}(\bR) & = & 2\sum_{n=1}^{N_s} \fint_{BZ} g_{n}(\bk) \lambda_{n}(\bk) \, \mathrm{d\bk} + E_{xc} (\rho(\bx)) - \int_{\Omega} V_{xc}(\rho(\bx))\rho(\bx)\, \mathrm{d\bx} \nonumber + \frac{1}{2} \int_{\Omega}\big(b(\bx,\bR)-\rho(\bx)\big) \phi(\bx,\bR) \, \mathrm{d\bx}  \nonumber  \\ 
& + & E_{sc}(\bR) + 2 k_B T \sum_{n=1}^{N_{s}} \fint_{BZ} \Big( g_{n}(\bk) \log g_{n}(\bk) + \big(1-g_{n}(\bk)\big) \log \big(1-g_{n}(\bk)\big) \Big) \, \mathrm{d\bk} \,, 
\end{eqnarray}
where $E_{xc}$ is the exchange-correlation energy, and $E_{sc}$ is the that incorporates the self energy and repulsive energy correction  associated with the pseudocharges \cite{suryanarayana2017sqdft}. The corresponding Hellman-Feynman force on the $I^{th}$ nucleus takes the form \cite{ghosh2017extended}  
\begin{eqnarray}\label{eqn:force}
\mathbf{f}_I & = &  \sum_{I'} \int_{\Omega} \nabla b_{I'}(\bx,\bR_{I'}) \phi(\bx,\bR) \, \mathrm{d\bx} +  \mathbf{f}_{sc,I} - 4 \sum_{n=1}^{N_s} \fint_{BZ} g_{n}(\bk) \sum_{lm} \gamma_{Il} \nonumber \\
 & \times &   \Re \Bigg[ \bigg( \int_{\Omega} \psi_{n}^*(\bx,\bk) \tilde{\chi}_{Ilm}(\bx,\bR_{I}) \, \mathrm{d\bx} \bigg) \, \bigg(\int_{\Omega} \nabla \psi_{n}(\bx,\bk) \tilde{\chi}_{Ilm}^*(\bx,\bR_{I}) \, \mathrm{d\bx} \bigg) \Bigg] \, \mathrm{d \bk}   \,,
\end{eqnarray} 
where $\mathbf{f}_{sc,I} = - \frac{\partial E_{sc}(\bR) }{ \partial \bR_I }$ and $\Re$[.] denotes the real part of the bracketed expression. As in previous work, the derivative on the nonlocal projectors (with respect to the atomic position) has been transferred to the orbitals (with respect to space) \cite{hirose2005first}. This strategy has been adopted since the orbitals are typically much smoother than the projectors, which enables more accurate forces with relatively minor eggbox effects \cite{andrade2015real,pratapa2015spectral,ghosh2017sparc}. Note that in the above description, the Laplacian and gradient operators are defined with respect to the standard Cartesian coordinate system, with derivatives expressed in terms of the lattice coordinates. 


\section{Kronecker product formulation of the real-space kinetic energy operator}\label{section:kroneckerproduct}
In terms of the lattice coordinates, the kinetic energy operator (i.e., Laplacian) for a non-orthogonal system takes the form:
\begin{equation}\label{eq:Laplacian}
\nabla^2 \equiv T_{11}\frac{\partial^2}{\partial\xi^2}+T_{22}\frac{\partial^2}{\partial\eta^2}+T_{33}\frac{\partial^2}{\partial\zeta^2}+(T_{12}+T_{21})\frac{\partial^2}{\partial\xi\partial\eta}+(T_{13}+T_{31})\frac{\partial^2}{\partial\xi\partial\zeta}+(T_{23}+T_{32})\frac{\partial^2}{\partial\eta\partial\zeta} \,,
\end{equation}
where $T_{ij}$ are the elements of the transformation matrix:
\begin{equation}
\mathbf{T} = 
\begin{bmatrix}
1 & \cos \gamma & \cos \alpha \\
\cos \gamma  &  1 & \cos \beta\\
\cos \alpha  &  \cos \beta &  1
\end{bmatrix}^{-1} \,,
\end{equation}
with $\alpha$, $\beta$, and $\gamma$ being the angles between the coordinate axes, as shown in Fig.~\ref{Fig:unitcell}. The Laplacian-vector multiplication is the dominant cost in central finite-difference based real-space DFT calculations, particularly for small to moderate sized systems ($\sim 1000$ atoms) where the cubic scaling bottleneck has still not manifested itself \cite{ghosh2017sparc,ghosh2017extended}. This is especially the case for non-orthogonal crystal systems, wherein the presence of mixed derivatives in the Laplacian (Eq.~\ref{eq:Laplacian}) makes its product with a vector scale quadratically with respect to the finite-difference order $n_o$, i.e., $\mathcal{O}(fn_o^2+3n_o+1)$ compared to $\mathcal{O}(3n_o+1)$ for orthogonal systems, where $f$ is the number of mixed derivatives in the Laplacian. Considering that high-order finite-differences are typically employed (e.g., $12^{th}$ order \cite{PARSEC,ghosh2017sparc}), the cost of real-space DFT calculations for non-orthogonal crystal systems is significantly larger than their orthogonal counterparts. In order to significantly bridge this gap, we now develop a Kronecker product formulation in which the Laplacian-vector multiplication scales linearly with the order of the finite-difference approximation. 

Consider a uniform discretization of $\Omega$ with a grid having a spacing of $h_1$, $h_2$, and $h_3$ along the $\bxi$, $\benta$, and $\bzeta$ directions, respectively, such that $L_1 = n_1 h_1$, $L_2 = n_2 h_2$, and $L_3 = n_3 h_3$ ($n_1$, $n_2$, $n_3$ $\in \mathbb{N}$, $\mathbb{N}: \text{set of all natural numbers}$). On the corresponding 1D grids, let the central finite-difference approximations of $\dfrac{\partial^2}{\partial\xi^2}$, $\dfrac{\partial^2}{\partial\eta^2}$, $\dfrac{\partial^2}{\partial\zeta^2}$, $\dfrac{\partial}{\partial\xi}$, $\dfrac{\partial}{\partial\eta}$, and $\dfrac{\partial}{\partial\zeta}$ with Bloch-periodic boundary conditions be denoted by $\bDchis \in \C^{n_1 \times n_1}$, $\bDetas \in \C^{n_2 \times n_2}$, $\bDzetas \in \C^{n_3 \times n_3}$, $\bDchi \in \C^{n_1 \times n_1}$, $\bDeta \in \C^{n_2 \times n_2}$, and $\bDzeta \in \C^{n_3 \times n_3}$, respectively. The product of the discrete Laplacian with a vector $\bX \in \C^{n_1 \times n_2 \times n_3}$ can then be written as
\begin{eqnarray}
{\nabla^2_h} \bX &=& \bigg(T_{11}\Big[\bIc \otimes \bIb \otimes \bDchis\Big] + T_{22}\Big[\bIc \otimes \bDetas \otimes \bIa\Big] + T_{33}\Big[\bDzetas \otimes \bIb \otimes \bIa\Big] \nonumber\\
 &+& (T_{12}+T_{21})\Big[\bIc \otimes \bDeta \otimes \bDchi\Big] + (T_{13}+T_{31})\Big[\bDzeta \otimes \bIb \otimes \bDchi\Big] \label{Eq:DiscLapVecProduct}\\
 &+& (T_{23}+T_{32})\Big[\bDzeta \otimes \bDeta \otimes \bIa\Big]\bigg)\bigg(vec_{n_3}(vec_{n_2}\bX)\bigg) \,, \nonumber
\end{eqnarray}
where $\otimes$ denotes the Kronecker product \cite{van2000ubiquitous,golub2012matrix}, $vec$ denotes the vectorization operator \cite{van2000ubiquitous,golub2012matrix}, and $\bIa \in \R^{n_1 \times n_2}$, $\bIb \in \R^{n_2 \times n_2}$, and $\bIc \in \R^{n_3 \times n_3}$ are identity matrices. It follows from this Kronecker product decomposition that the finite-difference coefficients for the second-order mixed derivatives can be obtained via the product of the coefficients for the associated first-order derivatives \cite{Iwata2010}. However, this relation does not provide any computational gain for the Laplacian-vector multiplication and the overall DFT calculation, which is the main focus of this work. In order to do so, we use Roth's relationship \cite{roth1934, Abadir2005,hugo2013} to simplify each of the terms in Eq.~\ref{Eq:DiscLapVecProduct} as follows:
\begin{eqnarray}\label{eqn:D_xx}
\bigg[\bIc \otimes \Big(\bIb \otimes \bDchis\Big)\bigg] \bigg(vec_{n_3}(vec_{n_2}\bX)\bigg) &=& vec_{n_3}\Bigg[\Big(\bIb \otimes \bDchis \Big) (vec_{n_2}\bX) \bIcT \Bigg] \nonumber\\
&=& vec_{n_3}\Bigg[vec_{n_2}\Bigg(\bigwedge_{1 \leqslant k \leqslant n_3}\Big(\bDchis \bX_k \bIbT\Big)\Bigg)\Bigg] \nonumber\\
&=& vec_{n_3}\Bigg[vec_{n_2}\Bigg(\bigwedge_{1 \leqslant k \leqslant n_3}\Big(\bDchis \bX_k\Big)\Bigg)\Bigg]\,,
\end{eqnarray}  
\begin{eqnarray}\label{eqn:D_yy}
\bigg[\bIc \otimes \Big(\bDetas \otimes \bIa\Big)\bigg] \bigg(vec_{n_3}(vec_{n_2}\bX)\bigg) &=& vec_{n_3}\Bigg[\Big(\bDetas \otimes \bIa \Big) (vec_{n_2}\bX) \bIcT \Bigg] \nonumber\\
&=& vec_{n_3}\Bigg[vec_{n_2}\Bigg(\bigwedge_{1 \leqslant k \leqslant n_3}\Big(\bIa \bX_k \bDetasT\Big)\Bigg)\Bigg] \nonumber\\
&=& vec_{n_3}\Bigg[vec_{n_2}\Bigg(\bigwedge_{1 \leqslant k \leqslant n_3}\Big(\bX_k\bDetasT\Big)\Bigg)\Bigg]\,,
\end{eqnarray}
\begin{eqnarray}\label{eqn:D_zz}
\bigg[\bDzetas \otimes \Big(\bIb \otimes \bIa\Big)\bigg] \bigg(vec_{n_3}(vec_{n_2}\bX)\bigg) &=& vec_{n_3}\Bigg[\Big(\bIb \otimes \bIa \Big) (vec_{n_2}\bX) \bDzetasT \Bigg] \nonumber\\
&=& vec_{n_3}\Bigg[vec_{n_2}\Bigg(\bigwedge_{1 \leqslant k \leqslant n_3}\Big(\bIa \bX_k \bIbT\Big)\Bigg)\bDzetasT \Bigg] \nonumber\\
&=& vec_{n_3}\Bigg[(vec_{n_2}\bX)\bDzetasT\Bigg]\,,
\end{eqnarray}
\begin{eqnarray}\label{eqn:D_xy}
\bigg[\bIc \otimes \Big(\bDeta \otimes \bDchi\Big)\bigg] \bigg(vec_{n_3}(vec_{n_2}\bX)\bigg) &=& vec_{n_3}\Bigg[\Big(\bDeta \otimes \bDchi \Big) (vec_{n_2}\bX) \bIcT \Bigg] \nonumber\\
&=& vec_{n_3}\Bigg[vec_{n_2}\Bigg(\bigwedge_{1 \leqslant k \leqslant n_3}\Big(\bDchi \bX_k \bDetaT\Big)\Bigg)\Bigg] \, ,
\end{eqnarray}
\begin{eqnarray}\label{eqn:D_xz}
\bigg[\bDzeta \otimes \Big(\bIb \otimes \bDchi\Big)\bigg] \bigg(vec_{n_3}(vec_{n_2}\bX)\bigg) &=& vec_{n_3}\Bigg[\Big(\bIb \otimes \bDchi \Big) (vec_{n_2}\bX) \bDzetaT \Bigg] \nonumber\\
&=& vec_{n_3}\Bigg[vec_{n_2}\Bigg(\bigwedge_{1 \leqslant k \leqslant n_3}\Big(\bDchi \bX_k \bIbT\Big)\Bigg)\bDzetaT\Bigg] \nonumber \\
&=& vec_{n_3}\Bigg[vec_{n_2}\Bigg(\bigwedge_{1 \leqslant k \leqslant n_3}\Big(\bDchi \bX_k \Big)\Bigg)\bDzetaT\Bigg] \, ,
\end{eqnarray} 
\begin{eqnarray}\label{eqn:D_yz}
\bigg[\bDzeta \otimes \Big(\bDeta \otimes \bIa\Big)\bigg] \bigg(vec_{n_3}(vec_{n_2}\bX)\bigg) &=& vec_{n_3}\bigg[\Big(\bDeta \otimes \bIa \Big) (vec_{n_2}\bX) \bDzetaT \bigg] \nonumber\\
&=& vec_{n_3}\Bigg[vec_{n_2}\Bigg(\bigwedge_{1 \leqslant k \leqslant n_3}\Big(\bIa \bX_k \bDetaT\Big)\Bigg)\bDzetaT\Bigg] \nonumber \\
&=& vec_{n_3}\Bigg[vec_{n_2}\Bigg(\bigwedge_{1 \leqslant k \leqslant n_3}\Big(\bX_k \bDetaT \Big)\Bigg)\bDzetaT\Bigg] \,,
\end{eqnarray}  
where $\bigwedge_{1 \leqslant k \leqslant n_3}$ is the loop operator \cite{hugo2013}, defined to be the matrix multiplication with each frontal slice $\bX_k \in \C^{{n_1}\times{n_2}}$ of $\bX$. Thereafter, utilizing Eqs~\ref{eqn:D_xx}, \ref{eqn:D_yy}, \ref{eqn:D_zz}, \ref{eqn:D_xy}, \ref{eqn:D_xz}, and \ref{eqn:D_yz}, the Laplacian-vector multiplication in Eq.~\ref{Eq:DiscLapVecProduct} can be simplified to take the form:
\begin{eqnarray}
{\nabla^2_h} \bX &=& vec_{n_3}\Bigg[vec_{n_2}\Bigg(\bigwedge_{1 \leqslant k \leqslant n_3}\Big(T_{11}\bDchis\bX_k + T_{22}\bX_k\bDetasT + (T_{12} + T_{21})\bDchi\bX_k\bDetaT\Big)\Bigg)  \nonumber \\ 
&+& (vec_{n_2}\bX)T_{33}\bDzetasT + vec_{n_2}\Bigg(\bigwedge_{1 \leqslant k \leqslant n_3}\Big((T_{13}+T_{31})\bDchi\bX_k + (T_{23} + T_{32}) \bX_k\bDetaT\Big)\Bigg)\bDzetaT\Bigg] \,.
\end{eqnarray} 
In doing so, the cost of the Laplacian-vector multiplication scales as $\mathcal{O}(2 n_0 f + 3n_o+3)$, which provides significant improvement in the efficiency relative to the original quadratic scaling of $\mathcal{O}(fn_o^2+3n_o+1)$. For example, the number of operations in the Laplacian-vector multiplication for a triclinic system ($f=3$) reduces by a factor of $\sim 4.2$ for the commonly employed $12^{th}$ order finite-difference approximation \cite{PARSEC,ghosh2017extended}. In addition to the computational speedup, there is a significant reduction in computer memory storage for implementations which store the Laplacian. Indeed, these improvements in speed and storage are also applicable to the generation of the pseudocharges and the solution of the electrostatic Poisson equation. Note that the above Kronecker product formulation does not change the accuracy of the underlying central finite-difference approximation, i.e., the discretization error still scales as $\mathcal{O}(h^{n_0})$, where $h$ is the effective mesh-size.

\section{Implementation and results} \label{Sec:ImplementationResults}
We implement the above described real-space framework for non-orthogonal crystal systems in M-SPARC, a MATLAB version of the real-space DFT code SPARC \cite{ghosh2017sparc,ghosh2017extended}. As part of the electrostatics, we assign the pseudocharges to the grid using the discrete Laplacian, similar to the strategy adopted for orthogonal systems \cite{Suryanarayana2014524,ghosh2014higher}. In addition, the linear system corresponding to the Poisson problem in Eq.~\ref{eqn:poisson} is solved using the Alternating Anderson-Richardson (AAR) method \cite{pratapa2016anderson,suryanarayana2016alternating}. The electronic ground-state is determined using the Chebyshev-filtered subspace iteration (CheFSI) method \cite{zhou2006parallel,zhou2006self} with acceleration provided by the restarted Periodic Pulay method \cite{pratapa2015restarted,banerjee2015periodic}. Geometry optimization is performed using the the Polak-Ribiere variant of non-linear conjugate gradients with a secant line search \cite{Shewchuk1994}. All integrations are performed using the trapezoidal rule, utilizing the Jacobian associated with the transformation to a non-orthogonal grid. We refer the reader to previous work of the authors in the context of orthogonal systems \cite{ghosh2017sparc,ghosh2017extended} for a detailed description of the underlying finite-difference formulation and implementation that has been adopted here.

In all simulations, we employ a twelfth-order accurate finite-difference discretization, norm-conserving Troullier-Martins pseudopotentials \cite{Troullier}, the Local Density Approximation (LDA) \cite{Kohn1965} with the Perdew-Wang parametrization \cite{perdew1992accurate} of the correlation energy calculated by Ceperley-Alder \cite{Ceperley1980}, a smearing of $\sigma = 0.001$ Ha, and the Monkhorst-Pack \cite{monkhorst} grid for integration over the Brillouin zone. As representative non-orthogonal systems, we consider (i) hexagonal close packed (hcp) magnesium and (ii) triclinic silicon obtained by homogeneously deforming a diamond cubic unit cell of silicon. Wherever suitable, we compare the results obtained with those by the plane-wave code ABINIT \cite{gonze2009abinit,ABINIT}, choosing plane-wave cutoffs of $18$ Ha and $40$ Ha for the magnesium and silicon systems, respectively. This results in highly accurate reference energy and forces that are converged to within $3\times 10^{-6}$ Ha/atom and $2\times 10^{-6}$ Ha/Bohr, respectively.  


\subsection{Convergence with discretization}\label{Subsec:Convergencewithdiscretization}
\begin{figure}[h!]
\centering
\subfloat[Energy]{\label{fig:energyConvergence}\includegraphics[keepaspectratio=true,width=0.42\textwidth]{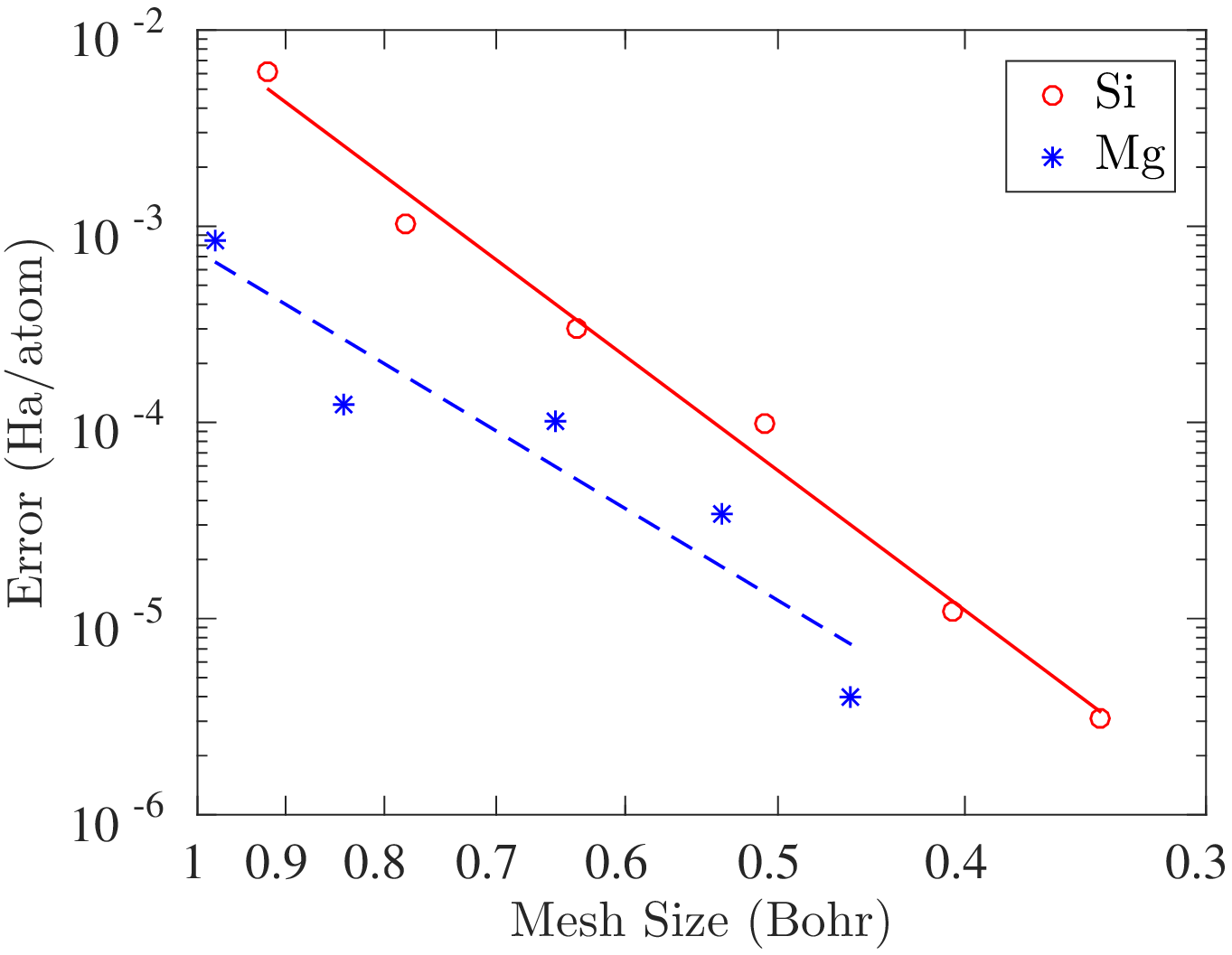}} \hspace{5mm}
\subfloat[Forces]{\label{fig:forceConvergence}\includegraphics[keepaspectratio=true,width=0.42\textwidth]{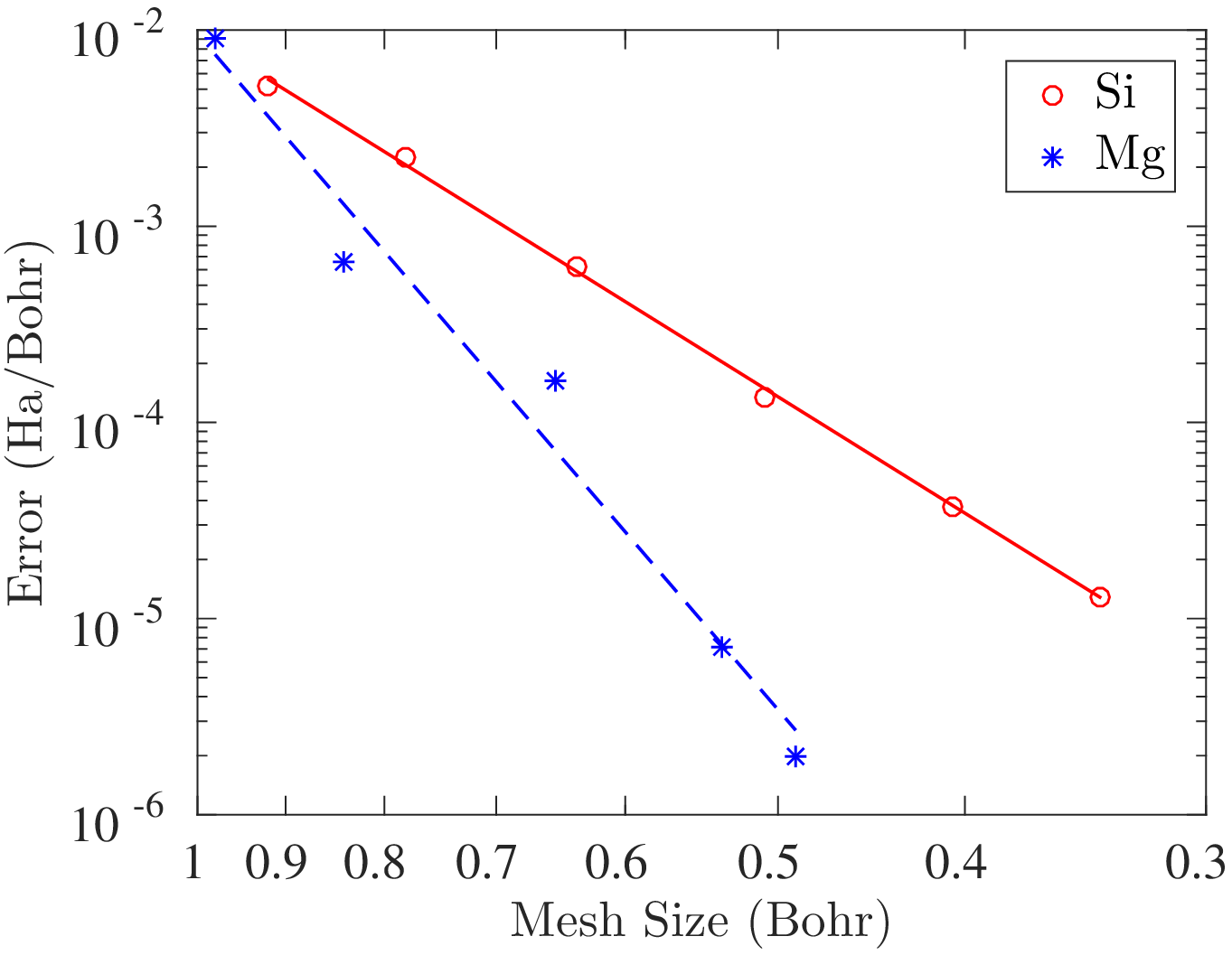}}
\caption{Convergence of the energy and atomic forces with respect to mesh size to reference planewave result for the hcp magnesium and triclinic silicon systems. The straight lines represent linear fits to the data.}
\label{fig:convergenceDiscretization}
\end{figure}

First, we verify convergence of the energy and atomic forces with respect to spatial discretization, all errors defined with respect to ABINIT. For this study, we consider: (i) a unit cell of hcp magnesium having lattice parameters: $L_1 = 5.87$ Bohr, $L_2 = 5.87$ Bohr, $L_3 = 9.62$ Bohr, $\alpha = 90^\circ$, $\beta = 90^\circ$, and $\gamma = 60^\circ$, with the interior atom perturbed by [$-0.14$ $0.12$ $2.1$] Bohr, and (ii) a unit cell of triclinic silicon having lattice parameters $L_1 = 10.16$ Bohr, $L_2 = 10.16$ Bohr, $L_3 = 10.16$ Bohr, $\alpha = 103^\circ$, $\beta = 82^\circ$, and $\gamma = 99^\circ$, with corner atom perturbed by [$0.51$ $0.41$ $0.31$] Bohr. It is clear from Fig.~\ref{fig:convergenceDiscretization}---plots of the error in energy and atomic forces with respect to mesh-size---that there is systematic convergence to the reference plane-wave result. On performing a fit to the data, we obtain average convergence rates of approximately $\mathcal{O}(h^{7})$ in the energy and $\mathcal{O}(h^{9})$ in the forces. These high convergence rates are similar to those obtained by SPARC for orthogonal systems \cite{ghosh2017extended}, thereby demonstrating the accuracy of the proposed formulation for non-orthogonal systems. Note that these numerically obtained convergence rates differ from the theoretical estimate associated with the discretization of the operators (i.e., $\mathcal{O}(h^{12})$ for twelfth-order accurate finite-differences). This difference can arise due to a number of factors, including the need for possibly finer meshes to obtain the asymptotic rates, nonlinear nature of the Kohn-Sham problem, and the use of trapezoidal rule for integration.

\subsection{Bulk properties}\label{Subsec:Bulkproperties}
\begin{figure}[h!]
\centering
\subfloat[Energy vs. volume]{\label{fig:EnergyLattice}\includegraphics[keepaspectratio=true,width=0.42\textwidth]{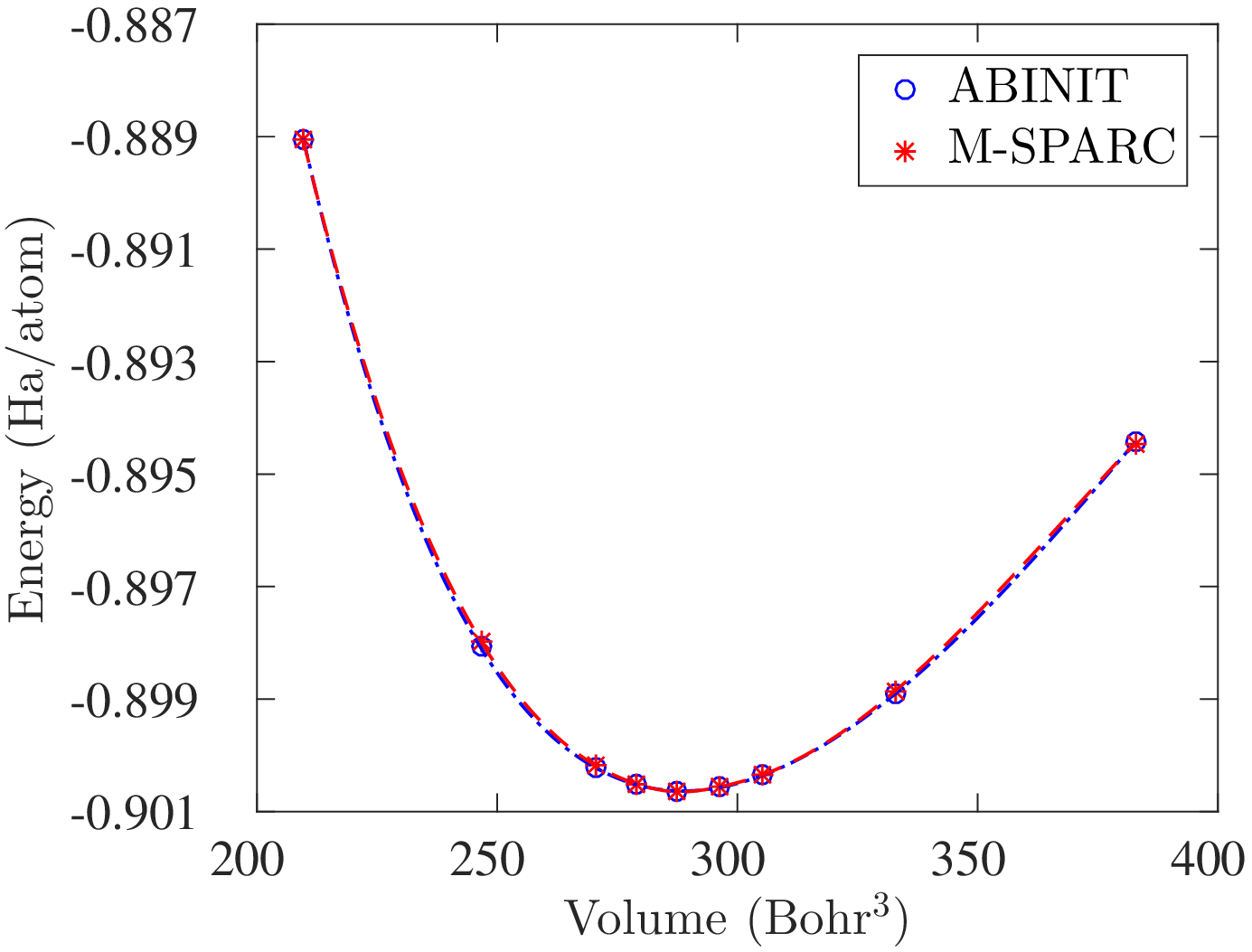}} \hspace{5mm}
\subfloat[Band structure diagram]{\label{fig:BandStructure}\includegraphics[keepaspectratio=true,width=0.42\textwidth]{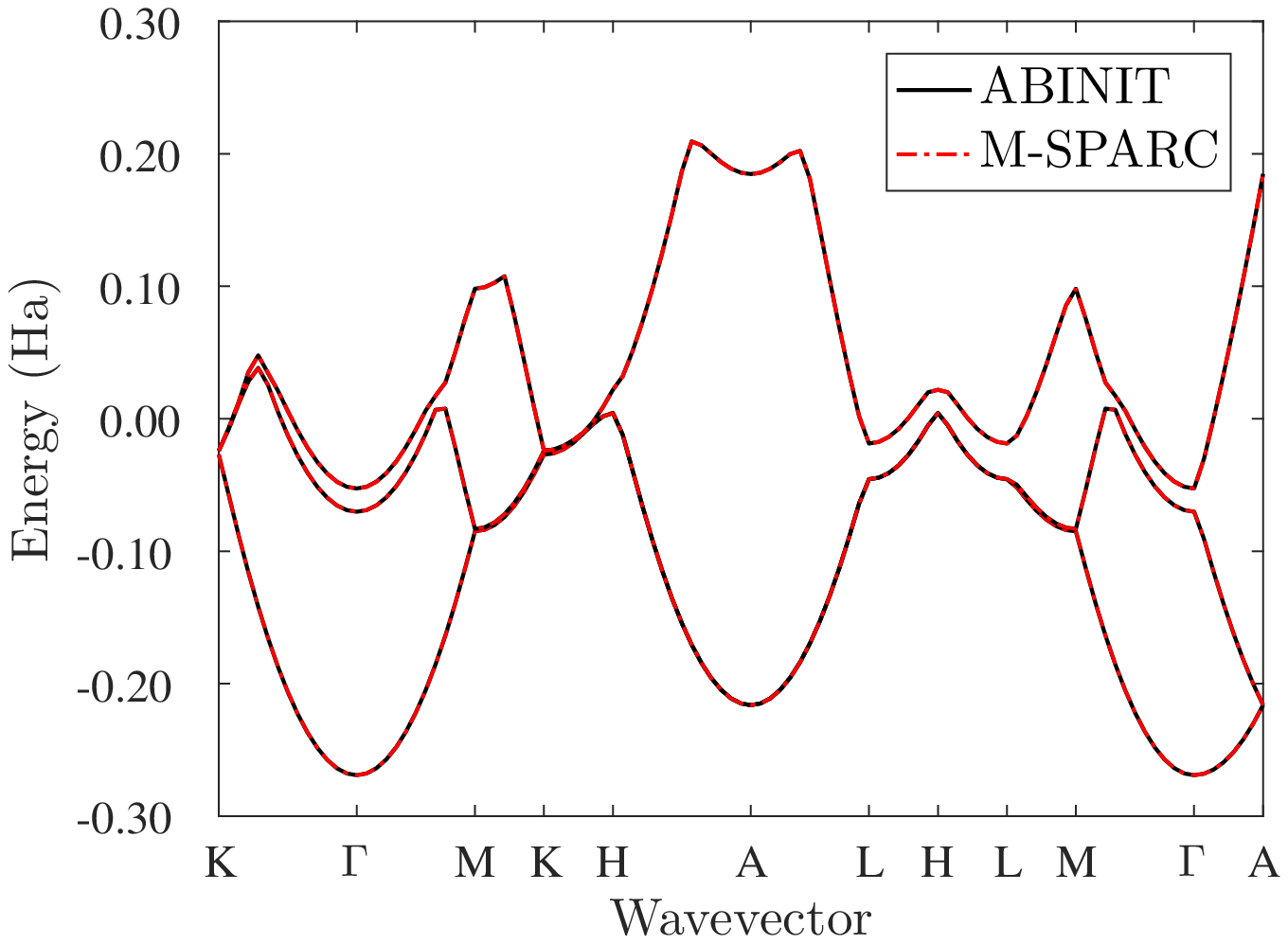}}
\caption{Bulk properties of hcp magnesium.}
\label{fig:bulkMg}
\end{figure}

We now verify the ability to accurately calculate bulk material properties, again using ABINIT results as reference. As the representative example, we consider a unit cell of hcp magnesium, with a mesh-size of $h = 0.65$ Bohr and $7\times 7\times 7$ grid for Brillouin zone integration. In Fig. \ref{fig:EnergyLattice}, we plot the energy so computed as a function of the volume of the unit cell. We observe that the curves are practically indistinguishable, demonstrating the excellent agreement with ABINIT results. Specifically, we find that the equilibrium lattice constant and energy---determined via a cubic spline fit to the data---are in agreement to within $0.006$ Bohr and $2 \times 10^{-5}$ Ha/atom, respectively.  At the equilibrium lattice constant so calculated, we compare the computed band structure diagram with ABINIT in Fig.~\ref{fig:BandStructure}. It is clear that the curves are nearly identical, again demonstrating the accuracy of the proposed real-space DFT formulation for non-orthogonal systems.

\subsection{Geometry optimization} \label{Subsec:Geometryoptimization}
\begin{figure}[h!]
\centering
\subfloat[Computed energy and its cubic spline fit]{\label{fig:EnergyDisplacement}\includegraphics[keepaspectratio=true,width=0.42\textwidth]{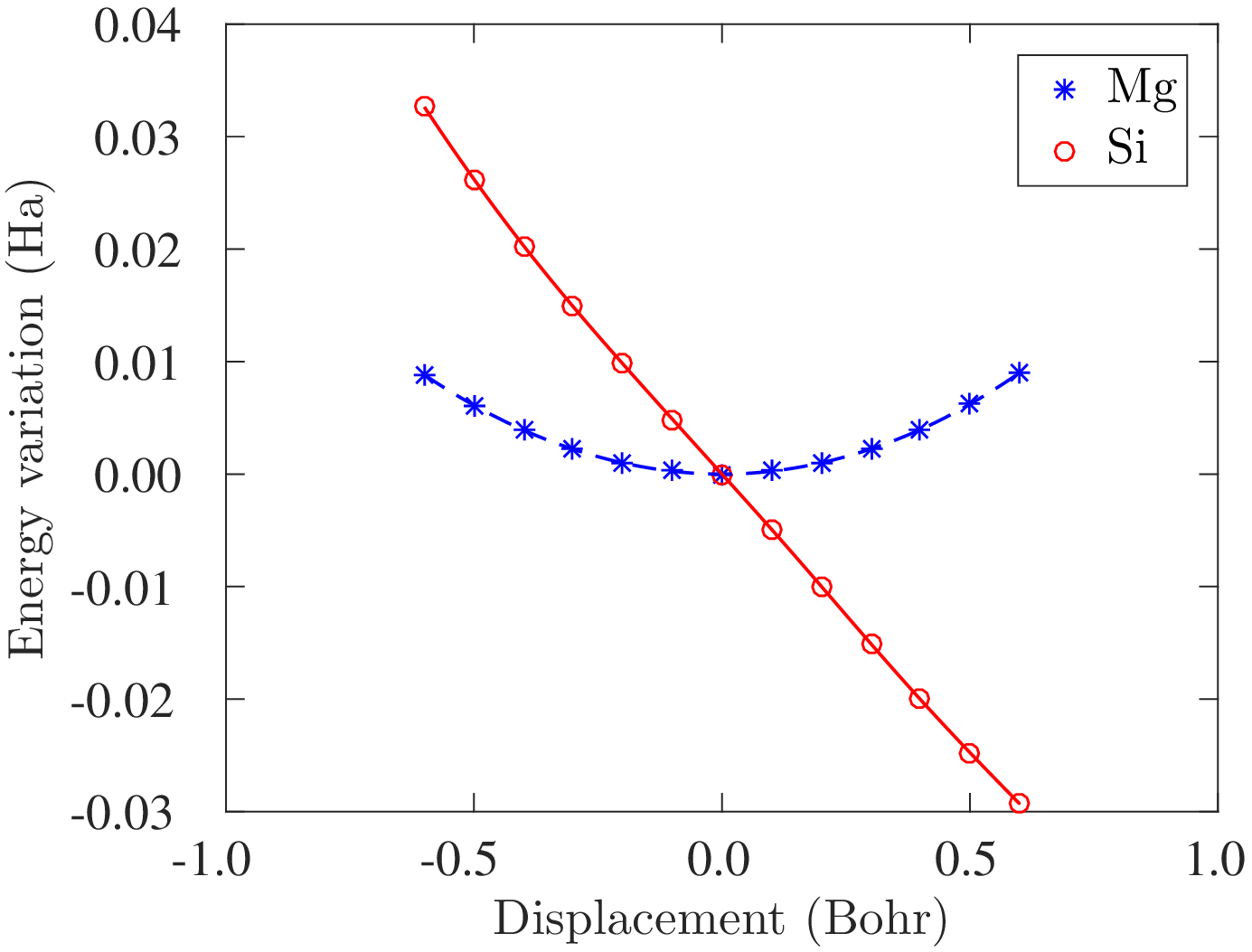}}
\subfloat[Computed force and the derivative of the cubic spline fit to the energy]{\label{fig:ConsistencyEnergyForce}\includegraphics[keepaspectratio=true,width=0.42\textwidth]{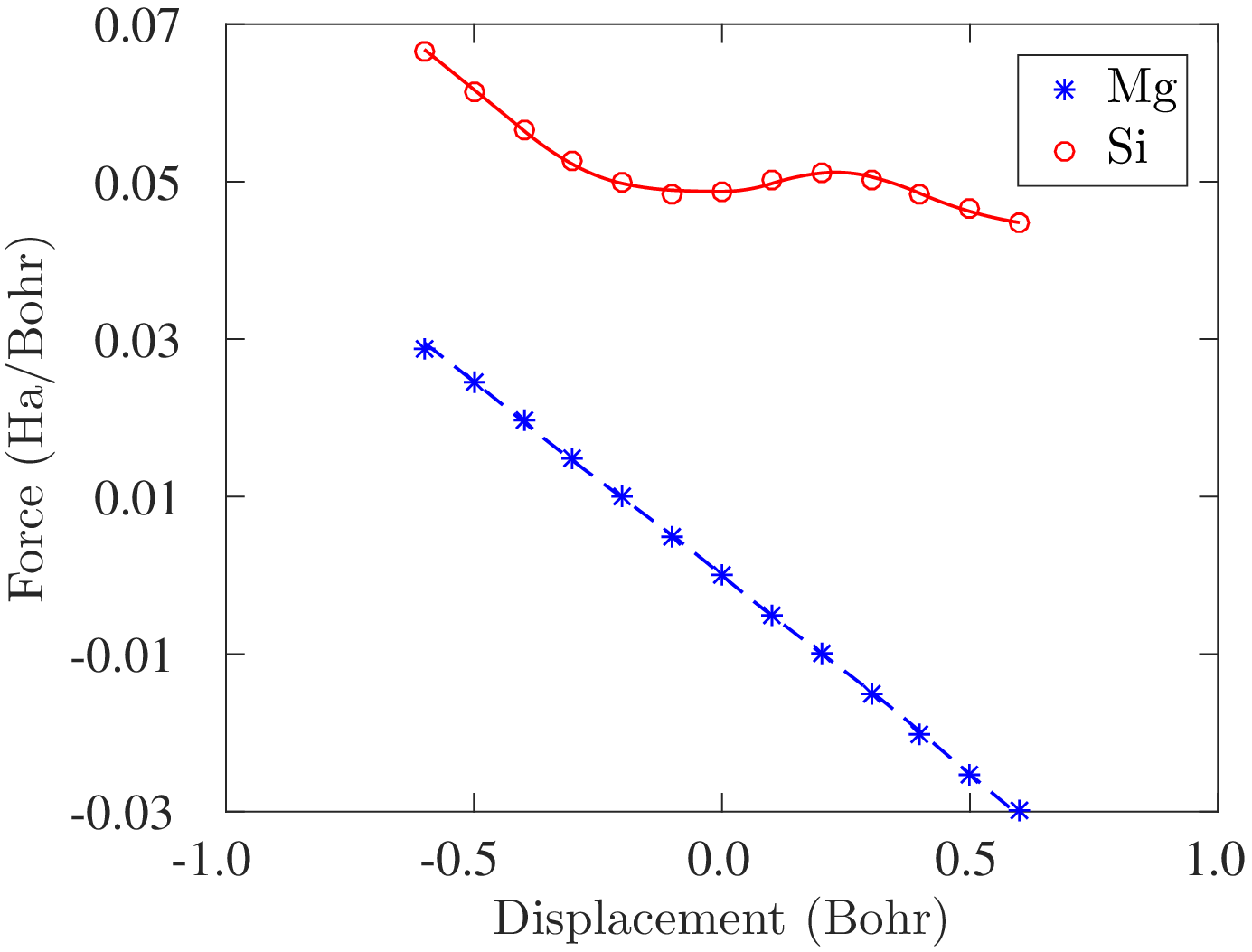}}
\caption{Variation in the energy and atomic force as a function of atomic displacement for the hcp magensium and triclinic silicon systems. The corner atoms are displaced along the cell diagonal and the cell edge in the magnesium and silicon systems, respectively.}
\label{Fig:EnergyForceConsistency}
\end{figure}

In order to verify the capability of the proposed framework to perform accurate geometry optimizations for non-orthogonal systems, we first check the consistency of the atomic forces with the energy. For this study, we consider unit cells of hcp magnesium and triclinic silicon---described in Section~\ref{Subsec:Convergencewithdiscretization}---for which we employ mesh-sizes of $h=0.65$ Bohr and $h=0.40$ Bohr, respectively. In Fig.~\ref{Fig:EnergyForceConsistency}, we plot the variation in energy and force when the corner atoms are displaced along the cell diagonal and cell edge in the magnesium and silicon systems, respectively. Specifically, we plot the computed energy and its cubic spline curve fit in Fig.~\ref{fig:EnergyDisplacement}. We plot the computed atomic force and the derivative of the curve fit to the energy in Fig.~\ref{fig:ConsistencyEnergyForce}. The excellent agreement demonstrates that the computed energy and atomic forces are consistent and that there is no noticeable `egg-box' effect \cite{brazdova2013atomistic}---a phenomenon arising due to the breaking of the translational symmetry---at meshes required for obtaining the accuracy desired in DFT calculations. 
Next, we determine the overall ground-state for $2 \times 2 \times 2$ unit cells of Mg with a vacancy. The computed vacancy formation energy \cite{gillan1989calculation,ghosh2017extended} is in agreement with ABINIT to within $4 \times 10^{-4}$ Ha and the fully relaxed atomic positions differ by no more than $1.8 \times 10^{-3}$ Bohr. 

\subsection{Performance} \label{Subsec:Performance}
Finally, we study the computational efficiency of the proposed Kronecker product formulation. We consider hcp magnesium and triclinic silicon systems of various sizes, with unit cells as described in Section~\ref{Subsec:Convergencewithdiscretization}. We employ mesh-sizes of $h=0.65$ Bohr and $h=0.40$ Bohr for the magnesium and silicon systems, respectively. In Table~\ref{Table:speedcomparison}, we compare the cost of the Laplacian-vector multiplication within the direct and Kronecker product implementations. We observe that the proposed approach is $\sim3$ and $\sim6$ times faster for the hcp and triclinic systems, respectively. These speedups are greater than the theoretically predicted ones---$\sim 2.9$ and $\sim 4.2$ for the hcp and triclinic systems, respectively---since the sparse matrices in the Kronecker product approach have a more compact banded structure compared to the sparse Laplacian matrix in direct multiplication. This translates to speedups of up to $\sim 1.3$ and $\sim 2.4$ in each SCF iteration for the hcp and triclinic systems, respectively. In addition, the cost relative to an analogous orthogonal system is only up to factors of $\sim 1.2$ and $\sim 1.3$ larger, respectively. 

\begin{table}[h!]
\centering
\begin{tabular}{|l|llll|llll|}
\hline
& Si$_8$ & Si$_{64}$ & Si$_{216}$ & Si$_{512}$
& Mg$_{16}$ & Mg$_{128}$ & Mg$_{432}$ & Mg$_{1024}$ \\\hline
Direct product & $0.08$ & $6.40$ & $73$ & $434$ & $0.025$ & $1.70$ & $21.0$ & $141$ \\
Kronecker product & $0.02$ & $1.06$ & $12$ & $79.0$ & $0.012$ & $0.63$ & $6.55$ & $40.0$ \\
\hline 
\end{tabular}
\caption{Computational time in seconds for a Laplacian-vector multiplication in the direct and Kronecker product methods.}
\label{Table:speedcomparison}
\end{table} 

As mentioned previously, an alternate technique to make the Laplacian-vector multiplication cost to scale linearly with the finite-difference order is to introduce additional directions (over which derivatives can be taken) in the Laplacian, removing all mixed derivatives in the process \cite{natan2008real}. However, in this approach, the effective grid spacing for the new directions can be significantly larger than that in the lattice vector directions, thereby limiting its accuracy/efficiency. For example, the mesh-size required by the technique of Natan et. al. \cite{natan2008real} for achieving an accuracy of $0.001$ Ha/atom (energy) and $0.001$ Ha/Bohr (forces) for the silicon systems described above is a factor of $\sim 1.2$ smaller than that required by the proposed approach. For the Si$_{512}$ system, this translates to the Kronecker product method being more efficient by factors of  $\sim 2.6$ and $\sim 2.0$ in the Laplacian-vector multiplication and SCF iteration, respectively. Indeed, these numbers are highly dependent on the geometry of the system. At the one end, for the hcp magnesium systems described above, the mesh-size required is nearly identical in both approaches, resulting in similar speeds.  At the other end, for a triclinic silicon system with $\alpha = 105^\circ$, $\beta = 75^\circ$, and $\gamma = 105^\circ$, the mesh-size required by the technique of Natan et. al. \cite{natan2008real} for achieving an accuracy of $0.001$ Ha/atom (energy) and $0.001$ Ha/Bohr (forces) is a factor of $\sim 1.6$ smaller. For the Si$_{256}$  system, this translates to the Kronecker product method being more efficient by factors of $\sim 6.8$ and $\sim 5.7$ in the Laplacian-vector multiplication and SCF iteration, respectively. Overall, these results demonstrate that the proposed real-space framework for non-orthogonal crystal systems is both accurate and efficient.


\section{Concluding Remarks} \label{Sec:conclusions}
In this work, we have presented a real-space framework for performing accurate and efficient Density Functional Theory (DFT) calculations of non-orthogonal crystal systems. Specifically, employing a local reformulation of the electrostatics that is equally applicable to systems with different crystal symmetries, we have developed a novel Kronecker product formulation of the real-space kinetic energy operator that significantly reduces the operation count associated with the Laplacian-vector multiplication, the dominant cost in real-space DFT simulations for small to moderate sized systems ($\sim 1000$ atoms). In particular, the scaling with respect to central finite-difference order is  reduced from quadratic to linear, thereby significantly bridging the gap in computational cost between non-orthogonal and orthogonal systems. We have demonstrated the accuracy and efficiency of the proposed methodology using hcp magnesium and triclinic silicon as representative examples. Overall, the proposed Kronecker product formulation of the kinetic energy operator overcomes one of the key limitations of real-space approaches, making them an even more attractive choice for DFT calculations.   
  

\section*{Acknowledgements}
The authors gratefully acknowledge the support of the National Science Foundation (CAREER - 1553212). The authors are also grateful to Qimen Xu for his help in writing the framework for M-SPARC.


\appendix

\section{Mathematical preliminaries}

\paragraph{Kronecker product ($\otimes$) \cite{van2000ubiquitous}}
The Kronecker product  of two matrices $ \bA \in \mathbb{C}^{n_1 \times n_2}$ and $ \bB \in \mathbb{C}^{n_3 \times n_4}$ resulting in a matrix $ \bC \in \mathbb{C}^{n_1n_3 \times n_2n_4}$ is represented as 
\begin{equation}
\bC = \bA \otimes \bB \,, \quad \text{where} \quad \bC(n_3(i-1)+p,n_4(j-1)+q) = \bA(i,j) \bB(p,q) \,.
\end{equation}

\paragraph{Vectorization operator ($vec$) \cite{golub2012matrix}}
The vectorization of a matrix $ \mathbf{A} \in \mathbb{C}^{n_1 \times n_2}$ resulting in a column vector $ \mathbf{B} \in \mathbb{C}^{n_1n_2 \times 1}$ is represented as
\begin{equation}
\mathbf{B} = vec_{n_2}(\mathbf{A})\,, \quad \text{where} \quad \bB((i-1)n_2+j,1) = \bA(i,j) \,.
\end{equation}

\paragraph{Loop operator ($\bw$) \cite{hugo2013}}
 The loop operator acting in the context of the product of matrices $ \bA \in \mathbb{C}^{n_1 \times n_2}$ and $ \bB \in \mathbb{C}^{n_2 \times n_3 \times n_4}$ resulting in a matrix $ \bC \in \mathbb{C}^{n_1 \times n_3 \times n_4}$ is represented as
\begin{equation}
\bw_{1 \leqslant k \leqslant n_4} \bA \bB_k =  \bC_k, \quad \text{where} \quad \bB_k = \bB(:,:,k) \text{ and } \bC_k = \bC(:,:,k) \,.
\end{equation}

\paragraph{Roth's relationship \cite{roth1934}}
 Given matrices $ \bA \in \mathbb{C}^{n_1 \times n_2}$, $ \bB \in \mathbb{C}^{n_2 \times n_3}$, and $ \bC \in \mathbb{C}^{n_3 \times n_4}$, it follows that
\begin{equation}
vec_{n_4}(\bA\bB\bC) = (\bC^T \otimes \bA)vec_{n_3}\bB \,.
\end{equation}

\bibliographystyle{ReferenceStyle}

\end{document}